# Nanoscale Field Effect Optical Modulators Based on Depletion of Epsilon-Near-Zero Films

Zhaolin Lu, Peichuan Yin, and Kaifeng Shi

*Microsystems Engineering PhD Program*
*Kate Gleason College of Engineering*
*Rochester Institute of Technology, Rochester, New York, 14623, USA*
zhaolin.lu@rit.edu

**Abstract**
   The field effect in metal-oxide-semiconductor (MOS) capacitors plays a key role in field-effect transistors (FETs), which are the fundamental building blocks of modern digital integrated circuits. Recent works show that the field effect can also be used to make optical/plasmonic modulators. In this paper, we report field effect electro-absorption modulators (FEOMs) each made of an ultrathin epsilon-near-zero (ENZ) film, as the active material, sandwiched in a silicon or plasmonic waveguide. Without a bias, the ENZ film maximizes the attenuation of the waveguides and the modulators work at the OFF state; contrariwise, depletion of the carriers in the ENZ film greatly reduces the attenuation and the modulators work at the ON state. The double capacitor gating scheme is used to enhance the modulation by the field effect. According to our simulation, extinction ratio up to 3.44 dB can be achieved in a 500-nm long Si waveguide with insertion loss only 0.71 dB (85.0%); extinction ratio up to 7.86 dB can be achieved in a 200-nm long plasmonic waveguide with insertion loss 1.11 dB (77.5%). The proposed modulators may find important applications in future on-chip or chip-to-chip optical interconnection.

## 1. Introduction

   With the successful demonstration of ultracompact lasers, detectors, and waveguides, the development of ultracompact optical modulators becomes one of the priority tasks for photonic integration. The existing techniques include electro-optical (EO) modulators based on linear EO effect in materials such as $LiNbO_3$ and free carrier plasma dispersion effect [1], as well as electro-absorption (EA) modulators based on Franz-Kyldish effect [2] and quantum-confined Stark effect [3]. However, they suffer either large dimensions or difficult integration.

   The most widely discussed EO modulators in academia these years are Si modulators based microrings [4-6]. The dimensions of Si modulators with high-Q microring can reduce down to a few micrometers. However, they suffer thermal instability. Even for the recently developed athermal modulators, they work well only within 7°C [7]. Seeing that thermal heaters are required to maintain the temperature on microring modulators, it is ironical that optical interconnection was initially proposed as a paradigm to decrease the power consumption of integrated circuits.

## 2. Theory of Operation

   Compared to the *pn* or *pin* structures in microring modulators, optical modulators may be also realized based on the MOS structure, which roughly functions as a parallel capacitor. When a bias voltage is applied across the oxide layer, charge is induced at metal and semiconductor surfaces. The field-induced charge per unit area in the MOS capacitor can be calculated according to

$$Q_s = \varepsilon E \approx \varepsilon \left(\frac{V}{d}\right), \qquad (1)$$

where $\varepsilon$ is the DC or RF permittivity of the insulator, and $E$ is the applied electric field across the insulator layer with thickness $d$. According to the bias polarity and strength, a MOS structure may work in three well-known modes: accumulation, depletion, and inversion. Generally speaking, accumulation induces more majority carriers, and the accumulation layer becomes more electrically conductive and more optically absorptive, whereas depletion removes free carriers, and the optical absorption by free carrier in the depletion layer can reduce to a negligible level. The inversion between electron and holes requires slow thermal excitation; thus, inversion is insignificant for high speed optical modulation. The accumulation layer thickness can be estimated as $l_{ac} \approx \pi L_D/\sqrt{2}$, where $L_D$ is the Debye length of the semiconductor. The depletion layer thickness can be estimated by

$$w_d = \frac{Q_s}{eN_0}. \qquad (2)$$

Dionne, *et al.* demonstrated an FEOM based on crystalline silicon sandwiched into two silver films [8], where the modulating electric field can switch the waveguide between guiding and cut-off states. Feigenbaun *et al.* investigated an FEOM based on conductive oxide (COx) [9], which is the first experimental demonstration of COx as active material for optical modulation. In particular, unity-order refractive index change in the COx accumulation layer was measured based on the ellipsometry method. The result is very promising and many works have followed this work, including numerical studies reported in Refs. [10-16]. Experimental demonstrations of COx as active material for EA/EO modulation are also reported [17-19]. Modulation extinct ratio is reported 1 dB/μm in Ref. [17], and 2.71 dB/μm in Ref. [18].

We are interested in EA modulation based on ENZ materials. Epsilon-near-zero materials [20-22] were initially proposed for microwave applications to improve the directivity of antennas [23], super coupling [24-26], and super tunneling [27], etc. We noticed that ENZ can also be achieved in graphene [28] and COx [11] for near-infrared applications. In particular, when a thin ENZ film is sandwiched in a single mode dielectric or plasmonic waveguide, very strong electric field can be excited in the ENZ film for the transverse magnetic (TM) mode. The structure was named as the "ENZ-slot" waveguide [11, 28，29], which is the extreme case for the slot waveguide [30]. The greatly enhanced electric field leads to greatly enhanced light absorption [11, 28].

Our previous work reported EA modulators based on ENZ-slot waveguides by accumulation [11]. However, the electron density in the accumulation layer nearly exponentially decays with distance, which leads to very non-uniform optical dielectric constant across the accumulation layer. Our recent experiment shows that both accumulation and depletion play an important role in the optical modulation [31, 32]. Herein, we propose EA modulators each based on the depletion of an ENZ thin film. In contrast to accumulation, the initial film can be designed with uniform carrier concentration and uniform ENZ optical dielectric constant. The depletion can greatly remove the carriers. Thus, optical modulation can be achieved between ENZ and depletion with very large extinction ratios. As examples, we consider EA modulators based on the metal-insulator-metal waveguide and Si waveguide platforms, respectively.

## 3. Design and Simulation

The challenge of this approach is that it requires high-k oxide and an ultrathin ENZ film. To overcome this, we employ the double capacitor gating scheme, as illustrated in Fig. 1(a), where the same gate voltage can induce depletion layers simultaneously on top and bottom sides of the ENZ film. The ultrathin film may be 2D semiconductor or COx. As an example, we first consider an optical modulator based on "metal-insulator-COx-insulator-metal" (MICIM), where a widely used COx, indium tin oxide (ITO) works at ENZ, and the metal layer is assumed to be Au owing its chemical stability and low absorption at near-infrared frequencies. Figure 1 (a) illustrates the MICIM structure to be investigated. Electro-absorption modulators based on the MICIM structure were investigated in our recent experiment [33]. According to the Drude model, the optical dielectric constant of COx can be approximated as

$$\epsilon_r = \epsilon' + j\epsilon'' = \epsilon_\infty - \frac{\omega_p^2}{\omega(\omega+j\gamma)} = \left[\epsilon_\infty - \frac{\omega_p^2}{\omega^2+\gamma^2}\right] + j\left[\frac{\omega_p^2 \gamma}{\omega(\omega^2+\gamma^2)}\right], \quad (3)$$

where $\epsilon_\infty$ is the high frequency dielectric constant, $\omega_p$ is the plasma frequency, and $\gamma$ is the electron damping factor. Plasma frequency, $\omega_p = \sqrt{\frac{Ne^2}{\epsilon_o m^*}}$, depends on carrier concentration $N$, and the effective electron mass $m^*$. Based on the typical optical parameters for ITO [10]:

$$\epsilon_\infty = 3.9, \quad m^* = 0.35 m_e, \quad \gamma = 1.8 \times 10^{14} \text{ rad/s},$$

ENZ can be achieved when $N = 6.32 \times 10^{20}$ cm$^{-3}$ at $\lambda_o = 1550$ nm. The corresponding optical dielectric constant is $\epsilon_r(enz) = 0.0942 + j0.563$, where the minimum $|\epsilon_r|$ can be achieved. Assume the insulator layers are both 10-nm thick HfO$_2$, which has a DC and RF dielectric constant of $\varepsilon_r = 25$ and dielectric strength > 1 V/nm. If 10-V positive bias is applied on the ITO, the depletion layer thickness can be estimated to be ~2.2 nm according to Eqs. (1 & 2). If the thickness of ITO is designed as 4 nm, then the entire ITO film can be mostly depleted by the double capacitor gating scheme. Figure 1(b) depicts the residual electron distribution for the depletion based on the finite element method (FEM) simulation, where a 10-V positive bias is applied at the left end of ITO film. Figure 1(c, d) is 1D scan of electron distribution in Fig. 1(b) at $x = 0$. As can be seen, the maximum electron density at the center ($y = 0$) in ITO reduces to $3.93 \times 10^{20}$ cm$^{-3}$ and then sharply decreases to a negligible level within 0.9 nm on both sides. The effect of induced surface charge on the optical properties of gold is negligible as Au has a huge free electron density, ~$5.9 \times 10^{22}$ cm$^{-3}$. Based on the Drude model, we calculated the complex optical dielectric constant distribution in the ITO as shown in Fig. 1 (e, f).

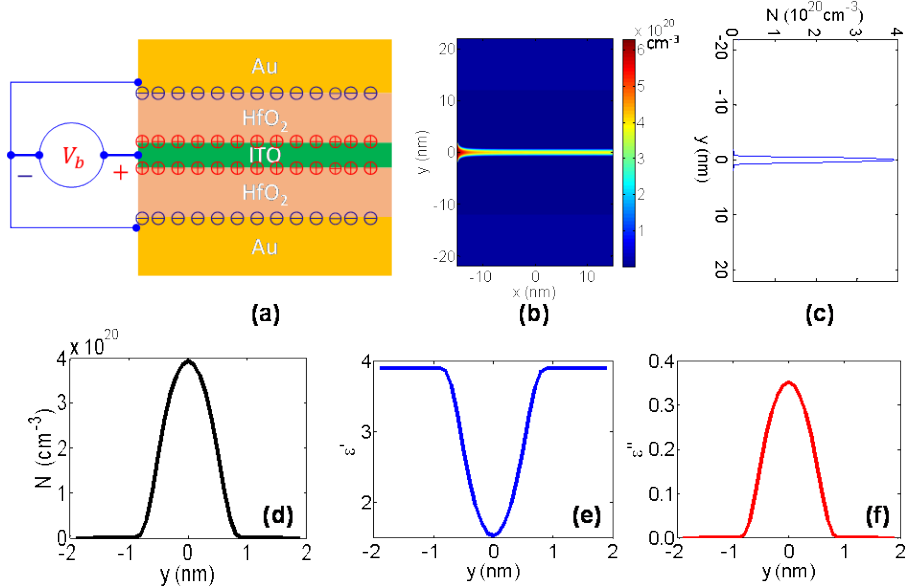

Figure 1: (a) Illustration of the metal-insulator-COx-insulator-metal structure working at the depletion mode. (b-d) FEM simulation of the residual electron distribution in the ITO film. (e) The real part of the ITO complex dielectric constant ($\epsilon'$) of ITO. (f) The imaginary part of the complex dielectric constant ($\epsilon''$) of ITO.

The double capacitor-gated ITO film can also be used in Si-based waveguides when the metal layers are replaced by $n^+$ Si, as illustrated in Fig. 2(a). In this case, the HfO$_2$ layers are each assumed as 10 nm thick, too. Note that the $n^+$ Si layers work at accumulation when the ITO works at depletion. Assume the carrier concentration of the $n^+$ Si is $N = 1 \times 10^{18}$ cm$^{-3}$ and the flat-band condition is satisfied without bias. The thickness of the accumulation layer in Si is less than 1 nm. Figure 2(b, c) depicts the electron density distribution in the ITO and Si. As shown in Fig. 2(d-f), the electron density and complex optical dielectric constant distributions in ITO are very similar to those in the plasmonic device as shown in Fig. 1(d-f). The maximum residual ITO electron density at the center ($y = 0$) is $4.58 \times 10^{20}$ cm$^{-3}$, which is slightly higher than the corresponding value in the plasmonic device, as shown in Fig. 2(a), because the potential drop in the Si accumulation layer reduces the voltage applied across the HfO$_2$ layers.

The accumulation electron distribution in Si is plotted in Fig. 2(g). The accumulated charge is distributed from above $2 \times 10^{21}$ cm$^{-3}$ at the surface decaying to $5 \times 10^{20}$ cm$^{-3}$ within 0.6 nm. The effect of the induced charge will result in index change, $\Delta n$, and absorption, $\alpha$, which can be estimated according to the following equations when working wavelength is $\lambda_o = 1550$ nm [1, 34].

$$\begin{cases} \Delta n = \Delta n_e + \Delta n_h = -[8.8 \times 10^{-22}\Delta N_e + 8.5 \times 10^{-18}(\Delta N_h)^{0.8}] \\ \Delta \alpha = \Delta \alpha_e + \Delta \alpha_h = 8.5 \times 10^{-18}\Delta N_e + 6.0 \times 10^{-18}\Delta N_h \end{cases}$$

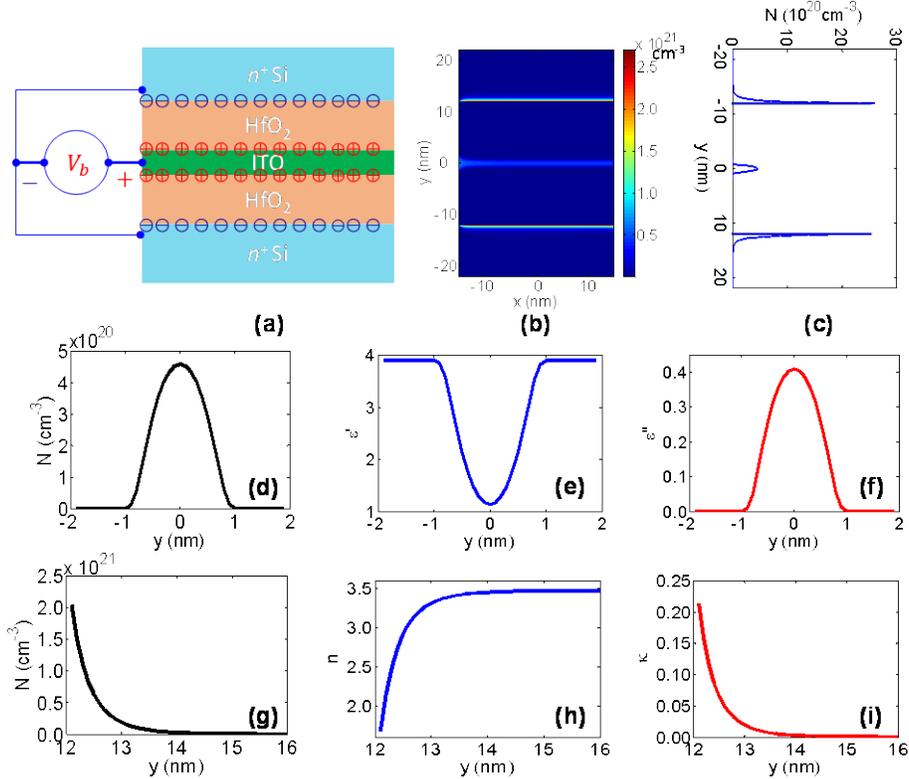

Figure 2: (a) Illustration of the Si-insulator-COx-insulator-Si structure working at the depletion mode. (b, c) FEM simulation of the residual electron distribution in the ITO film and accumulated electron distribution in Si. (d) The residual electron distribution in the ITO film. (e, f) The complex optical dielectric constant of ITO. (g) The accumulation electron distribution in the Si. (h, i) The complex refractive index of Si in the accumulation layer.

The corresponding complex refractive index of Si can be calculated according to

$$\begin{cases} n = n_o + \Delta n \\ \kappa = \dfrac{\alpha \lambda_o}{4\pi} \end{cases}. \quad (4)$$

Figure 2 (h, i) plots the complex refractive index, $n + j\kappa$, distribution of Si in the accumulation layer, where the refractive index of intrinsic Si $n_o = 3.476$ at $\lambda_o = 1550$ nm is assumed.

## 4. Performance Analysis

To consider the field effect on light propagation, we first put the Au-sandwiched structure in a 3D plasmonic waveguide as shown in Fig. 3(a). The inset illustrates the coordinates used in this work. The modulators to be investigated are much more sensitive to TM polarization (with major $E_y$ and $H_x$ components). The TM polarization and working wavelength $\lambda_o = 1550$ nm are assumed throughout this work. As can be seen, the top and bottom Au layers are electrically connected to easy apply the bias. Any bias applied between the ITO and the bottom Au layer will be automatically applied between ITO and the top Au layer with the same value.

We assume without bias the optical dielectric constant of is $\epsilon_r(enz) = 0.0942 + j0.563$. Using a 3D mode solver based on the finite-difference time-domain (FDTD) method, we found the mode supported by the plasmonic waveguide. Figure 3(b) shows the mode profile ($|E(x,y)|$) when the ENZ ITO is sandwiched. As can be seen, very strong electric field is excited in the ENZ ITO film, which results in a very large attenuation $\alpha = 59.27$ dB/μm. The effective index $n_{eff} = 1.840$ is considerably small because a large portion of power is concentrated in the low index ITO film. For the depletion case, we applied the complex dielectric constant of ITO as shown in Fig. 1(e, f) in the 3D mode solver. Figure 3(c) shows the corresponding mode profile, where the electric field and hence power distribution spread nearly the whole spacing between the two Au films. It is found that the attenuation reduces to $\alpha = 4.27$ dB/μm and effective index increases to $n_{eff} = 3.160$. In fact, most of the attenuation comes from the absorption by Au. The residual electrons in ITO only contribute ~0.5 dB/μm attenuation. In the ITO film layer, a mesh size $dy = 0.1$ nm is applied in the y-direction when the 3D mode solver is used. Outside the ITO layer, $dx = 2$ nm, $dy = 0.5$ nm.

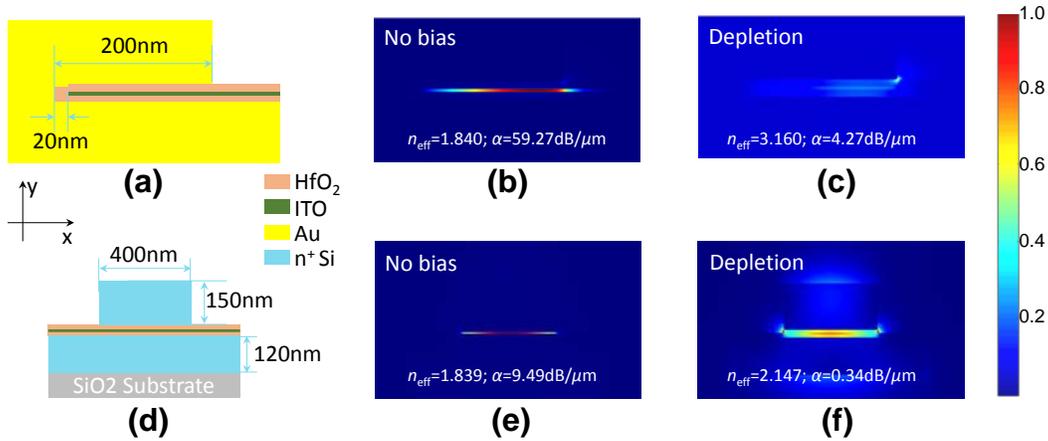

Figure 3. The electric field profiles ($|E(x,y)|$), effective indices, and propagation loss for different ITO-sandwiched waveguides in two cases: no bias and depletion. (a-c) In a plasmonic waveguide. (d-f) In a Si waveguide. The maximum field magnitude is normalized to be 1 in each plot. The refractive indices of Si, HfO$_2$, SiO$_2$, and Au are assumed to be 3.476, 2.070, 1.444, and 0.559+$j$9.81, respectively. The thickness of each HfO$_2$ layer is 10 nm, ITO thickness, 4 nm.

Similarly, we investigated the two operation modes when the ITO is sandwiched in a Si waveguide as shown in Fig. 3 (d). Figure 3(e) shows the mode profile ($|E(x,y)|$) when the ENZ ITO is sandwiched. Figure 3(f) shows the mode profile ($|E(x,y)|$) when the depleted ITO is sandwiched. The complex dielectric constant of ITO shown in Fig. 2(e, f) and refractive index of Si shown in Fig. 2 (h, i) are applied in the depletion case. As can be seen, the waveguide attenuation reduces from 9.49 dB/μm to 0.34 dB/μm and effective index slightly increases from 1.839 to 2.147 when ITO is switched from ENZ to depletion state. In the ITO film and Si accumulation layers, a mesh size $dy = 0.1$ nm is applied in the *y*-direction when the 3D mode solver is used. Elsewhere, $dx = 4$ nm, $dy = 0.5$ nm.

In order to evaluate the insertion loss of the EA modulators, we did 3D FDTD simulation with the mesh size $dx$ and $dy$ as described in above text and $dz = 2$ nm. We first simulated the modulator based on the plasmonic waveguide platform as shown in Fig. 4(a). In the design, the modulator is embedded in a waveguide with same configuration as itself except without the ITO layer. As a result, the input and output waveguides are two identical Au-HfO$_2$-Au waveguides, where the thickness of HfO$_2$ is 20 nm. The length of the EA modulator is 200 nm. Figures 4(b & c) show the power distribution in the waveguide in ENZ and depletion cases, respectively. Two detectors are placed 30 nm away from the modulator to evaluate the input and output power flow in the *z*-direction. Simulation results demonstrate that the overall throughput is 12.7% for ENZ ITO, and 77.5% for depleted ITO. Note that the insertion loss is 1.11 dB (77.5%). The achievable extinction ratio is 7.86 dB.

We also simulated the modulator based on the Si waveguide platform as shown in Fig. 4(d). The length of the EA modulator is 500 nm. In this case, the modulator is embedded in a Si waveguide with same overall dimensions as itself except without the ITO and HfO$_2$ layers. Figures 4(e & f) show the power distribution in the waveguide in ENZ and depletion cases, respectively. Simulation results demonstrate that the overall throughput is 38.5% for ENZ ITO, and 85.0% for depleted ITO, which give rise to extinction ratio, 3.44 dB with insertion loss, 0.71 dB.

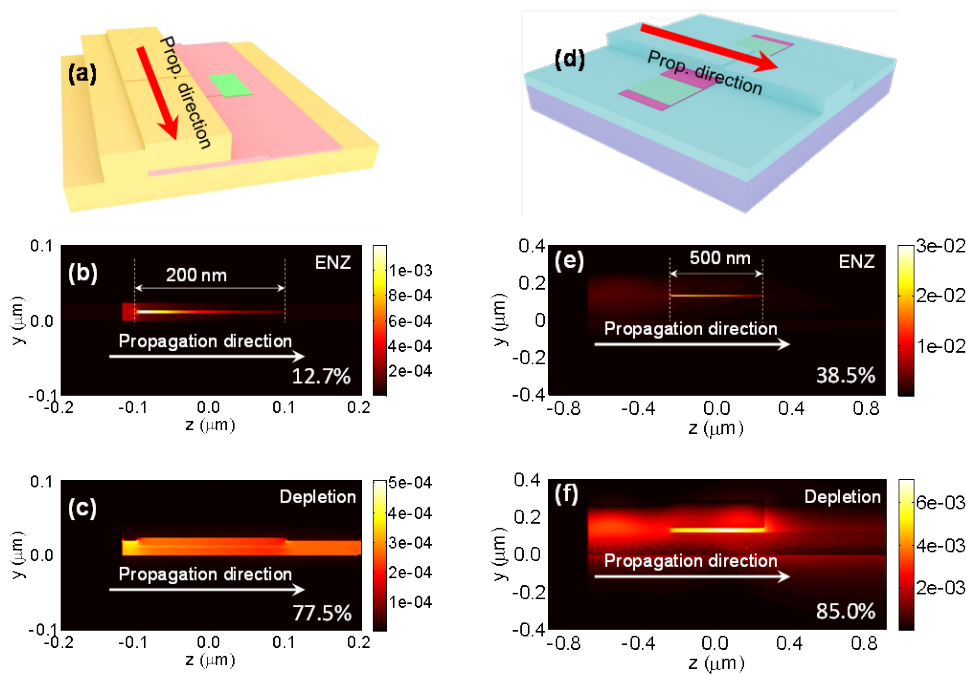

Figure 4. (a) The illustration of an EA modulator embedded in a Au-HfO$_2$-Au plasmonic waveguide. The cross section dimensions of the modulator are shown in Fig. 3(a). (b, c) The 3D simulation of light propagation between the plasmonic waveguide and the EA modulator in ENZ and depletion cases, respectively. (d) The illustration of an EA modulator embedded on a rib Si waveguide. The cross section dimensions of the modulator are shown as Fig. 3(d). (e, f) The 3D simulation of light propagation between the rib Si waveguide and the EA modulator in ENZ and depletion cases, respectively.

## 5. Discussion and Conclusion

The optical bandwidth of the modulators can be over several THz due to the slow Drude dispersion [11]. The EA modulators can potentially work at an ultra-high speed, being mainly limited by the RC delay imposed by electric circuits. In spite of the involvement of double, thin high-k insulator layer, the parasitic capacitance of the EA modulators can still be very small owing to their nanoscale footprints. Thus, low energy consumption per bit is expected. The fabrication of 4-nm thick ITO may be challenging. There are two approaches to address this issue:

(1) Multilayer 2D semiconductor with heavily doped carrier concentration. For example, 10 layers of graphene.
(2) Material with even higher dielectric concentration. There were reports on perovskite-based dielectrics, $BaTiO_3$ with $\varepsilon_r=165$ [35], and $SrTiO_3$ with $\varepsilon_r=219$ [36] at room temperature.

The decrease of the ENZ film or insulator thickness and the increase of the dielectric constant of insulator can further improve the performance of the proposed EA modulators.

To summarize, recent experiments show the potential applications of COx as tunable ENZ material. When sandwiched in a plasmonic or Si waveguide, a very thin ENZ film can greatly enhance light absorption; whereas, the depletion of the carriers in the ENZ film can significantly reduce light absorption. The depletion of ultrathin ENZ film-sandwiched waveguides may enable EA modulation at the nanoscale. The successful development of this technique may lead to a significant breakthrough in on-chip optical interconnects.

This material is based upon work supported by the National Science Foundation under Award No. ECCS-1308197.